\documentclass[11pt,notoc]{JHEP3}
\usepackage[fleqn]{amsmath}
\usepackage{bbm,bm}

\def\f{\phi}
\def\ts#1{{\textstyle #1}}
\def\tsfrac#1#2{\ts{\frac{#1}{#2}}}

\def\+{{+\!\!\!+}}
\def\-{=}

\def\d{{\rm d}}

\def\ts#1{{\textstyle #1}}
\def\-{=}

\def\g{\gamma}
\def\s{\sigma}

\def\gen#1{\mathcal{#1}}
\def\genJ{\gen{J}}

\ifx\mathbbm\undefined
  
\else
  
\fi
\ifx\bm\undefined

\else 

\fi

\def\matrix#1#2{\left(\begin{array}{#1}#2\end{array}\right)}
\def\gcgMatrix#1{\matrix{cc}{#1}}

\title{T-duality and Generalized Complex Geometry}

\author{
  \parbox{14cm}{
    Jonas Persson\footnotemark[1]
    }\\
  \parbox[t]{6.5cm}{
   \footnotemark[1]\,\,\parbox[t]{6cm}{
     Department of Theoretical Physics\\
     Uppsala University\\
     Box 803, SE-751\,08 Uppsala\\
     Sweden
   }
  }
  \ \\ \\
  {\tt Jonas.Persson@teorfys.uu.se
  }
}

\abstract{We find the explicit T-duality transformation in the phase space formulation of the $N=(1,1)$ sigma model. We also show that the T-duality transformation is a symplectomorphism and it is an element of $O(d,d)$. Further, we find the explicit T-duality transformation of a generalized complex structure in this model. We also show that the extended supersymmetry of the sigma model is preserved under the T-duality.
}

\preprint{
  hep-th/0612034\\
  UUITP-19/06
}

\begin{document}
\section{Introduction}
T-duality is an equivalence between physical observables in string theory on a space-time geometry and observables in string theory on a, possibly different, dual space-time geometry. This equivalence holds to all orders in string perturbation theory. The exact relations between the dual space-times, for the case where they are non-flat, are the so called Buscher rules, derived in \cite{Buscher:1987sk, Buscher:1987qj}. Classical T-duality of a sigma model can be described as a gauging, and subsequent gaugefixing, of an isometry of the target space \cite{Rocek:1991ps, Hull:1985pq}, or as a canonical transformation of the corresponding phase space formulation \cite{Curtright:1994be, Alvarez:1994wj, Sfetsos:1996pm}. T-duality has been studied extensively over the years, for details see the reviews \cite{Alvarez-Gaume:1993nk, Giveon:1994fu, Alvarez:1994dn} and references therein. For an introduction to dualities see e.g.\ \cite{Hjelmeland:1997eg}.

In this paper we focus on the interpretation of T-duality as a canonical transformation and study the simplest case, where the target space has one abelian isometry. One question we want to answer is whether it is possible to find the T-duality transformation in the phase space formulation of the $N=(1,1)$ supersymmetric sigma model. This question was also addressed in \cite{Hassan:1995je}, the manifest transformation given therein is only valid on shell, hiding the transformation of the conjugate momenta. Here we study the phase space formulation of the $N=(1,1)$ supersymmetric sigma model \cite{Zabzine:2005qf}. We find the T-duality transformation, valid off-shell, for the $N=1$ phase space superfields and give it explicitly. We also find that it is a symplectomorphism.

It is known since \cite{Zumino:1979et} that supersymmetry and target space geometry has a deep relation. The concept of Generalized Complex Geometry was introduced in \cite{Hitchin:2004ut} and later developed in \cite{Gualtieri:2003dx}. T-duality in Generalized Complex Geometry has previously been addressed in \cite{Cavalcanti:2005hq, Ben-Bassat:2004vn}. The connection between Generalized Complex Geometry and supersymmetric sigma models has been extensively studied \cite{Bergamin:2004sk, Bredthauer:2005zx, Calvo:2005ww, Lindstrom:2004eh, Lindstrom:2004iw}. For reviews see \cite{Zabzine:2006uz, Lindstrom:2006ee}. For example, in the phase space formulation \cite{Zabzine:2005qf} of the $N=(1,1)$ sigma model, extending the supersymmetry to $N=(2,2)$ requires the target space to be generalized K\"{a}hler \cite{Bredthauer:2006hf} and to $N=(4,4)$ requires the target space to be generalized Hyperk\"{a}hler \cite{Bredthauer:2006sz, Ezhuthachan:2006yy}.

Whether or not the extended supersymmetry survives the T-duality transformation can be studied using Generalized Complex Geometry. In this paper we study the $N=(1,1)$ sigma model and its extended supersymmetry. We assume that the target space has an isometry, the direction in which we T-dualize. We find the explicit T-duality transformation of a generalized complex structure and find that its T-dual also is a generalized complex structure, implying that the extended supersymmetry survives the transformation. Hence, we explore the results of \cite{Cavalcanti:2005hq} in a sigma model framework. To have extended supersymmetry the model has to have a generalized K\"{a}hler or generalized Hyperk\"{a}hler target space, and thus it should be emphasized that we in this paper study the T-dual of generalized K\"{a}hler and of generalized Hyperk\"{a}hler geometry.

The paper is organized as follows. In Section 2 we briefly review two dimensional superspace and introduce the $N=(1,1)$ supersymmetric sigma model. In Section 3 we describe how to find the phase space formulation of this model. In Section 4 we study the gauging procedure that give the T-dual model. In Section 5 the T-duality transformation for the phase space formulation is derived. We also show that it is a symplectomorphism. In Section 6 we formulate the T-duality transformation in terms of matrices acting on sections of $TM\oplus T^*M$, we also find the transformation of a generalized complex structure in this model. Section 7 is devoted to proving the integrability of the T-dual generalized complex structure. In section 8 we comment on different amount of extended supersymmetry. In Section 9 we give two explicit examples of T-dual spaces and in Section 10 a summary and open questions.

\section{The $N=(1,1)$ supersymmetric sigma model}
In this section we state the relevant properties of two dimensional superspace and  formulate the $N=(1,1)$ sigma model on this space. For more details see \cite{Bredthauer:2006hf}.

In two-dimensional superspace we denote the two spinor coordinates 
by $\theta^\pm$, the spinor derivatives by $D_\pm$ and the supersymmetry generators by $Q_\pm$. 
They obey the relations
\begin{align}
D_\pm^2 = i (\partial_0 \pm \partial_1), \;\;\;\;\;\;\;
\{D_+,D_-\} = 0, \;\;\;\;\;\;\;
Q_\pm = i D_\pm + 2\theta^\pm (\partial_0 \pm \partial_1). \label{eq:spinor_relations}
\end{align}
The supersymmetry transformation of a superfield $\Phi$ is given by
\begin{align}
\delta\Phi^\mu = -i (\epsilon^+ Q_+ + \epsilon^- Q_-)\Phi^\mu.
\end{align}

We now formulate the manifest $N=(1,1)$ sigma model in two-dimensional superspace and review the results of \cite{Gates:1984nk}. The action is given by 
\begin{align}
S= \frac{1}{2} \int\d^2\s\d\theta^+\d\theta^- \; D_+ \Phi^\mu D_-\Phi^\nu E_{\mu\nu}(\Phi),\label{eq:N11action}
\end{align}
where $E_{\mu\nu}(\Phi)=g_{\mu\nu}(\Phi)+b_{\mu\nu}(\Phi)$. This action is invariant under the additional supersymmetry transformation
\begin{align}
\delta\Phi^\mu = \epsilon^+ J^\mu_{(+)\nu}D_+\Phi_\nu + \epsilon^- J^\mu_{(-)\nu}D_-\Phi^\nu \label{eq:N22transfn}
\end{align}
if $J_{(\pm)}$ are two complex structures that satisfy $J^t_{(\pm)} g J_{(\pm)} = g$ and $\nabla^{(\pm)} J_{(\pm)} = 0$, where $\nabla^{(\pm)}$ is the covariant derivatives with the torsionful connections $\Gamma^{(\pm)} = \Gamma^{(0)} \pm g^{-1}H$. The Levi-Civit\'a connection  is denoted by $\Gamma^{(0)}$ and $H$ is the 3-form field strength of $b$, $H=db$. This geometry is called bi-Hermitean. Since $\epsilon^\pm$ are two independent parameters the transformation \eqref{eq:N22transfn} define one additional left going and one additional right going supersymmetry. Thus, the sigma model has $N=(2,2)$ supersymmetry if the target space is bi-Hermitean \cite{Gates:1984nk}.

\section{Phase space formulation of the sigma model}\label{sec:to_phase_space}
Since we eventually want to find the T-duality transformation in the phase space formulation of the above sigma model \eqref{eq:N11action}, we now review this phase space formulation. We follow the lines presented in \cite{Bredthauer:2006hf}.

To find the appropriate phase space formulation we introduce the new odd coordinates and spinor derivatives as
\begin{align}
&\theta^0 = \tsfrac{1}{\sqrt{2}}\left(\theta^+ - i \theta^-\right), \;\;\;\;
\theta^1 = \tsfrac{1}{\sqrt{2}}\left(\theta^+ + i \theta^-\right), \cr
&D_0 = \tsfrac{1}{\sqrt{2}}\left(D_+ + i D_-\right), \;\;\;\;
D_1 = \tsfrac{1}{\sqrt{2}}\left(D_+ - i D_-\right).\label{eq:newcoords}
\end{align}
The derivatives satisfy the algebra $D^2_0 = i\partial_1$, $D^2_1 = i\partial_1$, $\{D_0,D_1\} = 2i\partial_0$.

We want to perform the $\theta^0$ integration in the action \eqref{eq:N11action}
and for this we introduce the $N=1$ superfields
\begin{align}
\f^\mu = \Phi^\mu|_{\theta^0=0}, \;\;\;\;\;\;\;\;
S_\mu = g_{\mu\nu} D_0\Phi^\nu|_{\theta^0=0}. \label{eq:N=1superfields}
\end{align}
We also define $D\equiv D_1|_{\theta^0=0}$, $\theta\equiv\theta^1$, $\s \equiv \s^1$ and $\partial \equiv \partial_1$.

Changing the action \eqref{eq:N11action} to the new coordinates and performing the integration over $\theta^0$, by use of $\int\d\theta^0(\cdot) = D_0(\cdot)|_{\theta^0=0}$, gives the phase space action
\begin{align}
S = \int\d\s^0 \left(
        \left\{\int\d\s\d\theta \; i(S_\mu-b_{\mu\nu}D\f^\nu)\partial_0\f^\mu \right\}
          - \mathcal{H}\right), \label{eq:N1reducedaction}
\end{align}
where we identify the Hamiltonian as
\begin{align}
\mathcal{H}=
\frac{1}{2}\int\d\s\d\theta \;\Big(
&   D^2\f^\mu D\f^\nu g_{\mu\nu}
  + S_\mu DS_\nu g^{\mu\nu}
  + S_\mu D\f^\rho S_\nu g^{\nu\sigma}\Gamma^{(0)\mu}_{\sigma\rho}\cr
& + D\f^\mu D\f^\nu S_\rho H_{\mu\nu}{}^\rho
  -\frac{1}{3} S_\mu S_\nu S_\rho H^{\mu\nu\rho} \Big). \label{eq:Hamiltonian}
\end{align}
Here, $\Gamma^{(0)\mu}_{\sigma\rho}$ is the Levi-Civit\'a connection and $H_{\mu\nu\rho}$ is the 3-form field strength of $b_{\mu\nu}$ defined by $H_{\mu\nu\rho}=\frac{1}{2}\left(b_{\mu\nu,\rho}+b_{\nu\rho,\mu}+b_{\rho\mu,\nu}\right)$.

The first term in \eqref{eq:N1reducedaction} is the Liouville form $\Theta$ that defines the symplectic structure $\omega=\delta\Theta$, which in turn gives the Poisson bracket as \cite{Bredthauer:2006hf}
\begin{align}
\{F,G\} = i \int \d\s\d\theta \left(
  \frac{F \overleftarrow{\delta}}{\delta S_\mu}\frac{\overrightarrow{\delta}G}{\delta\f^\mu}
- \frac{F \overleftarrow{\delta}}{\delta\f^\mu}\frac{\overrightarrow{\delta}G}{\delta S_\mu}
+ 2 \frac{F \overleftarrow{\delta}}{\delta S_\nu} H_{\mu\nu\rho}D\f^\mu \frac{\overrightarrow{\delta} G}{\delta S_\rho}
\right). \label{eq:twisted_Poisson_bracket}
\end{align}
The functional derivatives are defined by 
\begin{align}
\delta F = \int\d\s\d\theta \left(
   \frac{F \overleftarrow{\delta}}{\delta S_\mu} \delta S_\mu 
+  \frac{F \overleftarrow{\delta}}{\delta\f^\mu} \delta\f^\mu
\right)
=\int\d\s\d\theta \left(
  \delta S_\mu \frac{\overrightarrow{\delta}F}{\delta S_\mu} 
   + \delta\f^\mu \frac{\overrightarrow{\delta} F}{\delta\f^\mu}
\right).\label{eq:def_func_deriv}
\end{align}

The Hamiltonian \eqref{eq:Hamiltonian} is invariant under the manifest supersymmetry transformation \cite{Bredthauer:2006hf}
\begin{align}
\delta_1 \f^\mu = -i \epsilon Q\f^\mu, \;\;\;\;\;\;\;\;
\delta_1 S_\mu = -i \epsilon Q S_\mu,
\end{align}
and under the non-manifest supersymmetry transformation 
\begin{align}
\tilde{\delta}_1 \f^\mu = \epsilon g^{\mu\nu} S_\nu, \;\;\;\;\;\;\;\;
\tilde{\delta}_1 S_\mu = i \epsilon g_{\mu\nu} \partial \f^\nu 
                         + \epsilon S_\lambda S_\sigma g^{\lambda\rho} \Gamma^\sigma_{\mu\rho}.
\end{align}
Further, the two generators
\begin{align}
\mathcal{Q}^{(i)}_2(\epsilon) = -\frac{1}{2}\int\d\s\d\theta\; \epsilon \left(
  2D\f^\mu S_\nu J^\nu_{(i)\mu} + D\f^\mu D\f^\nu L_{(i)\mu\nu} + S_\mu S_\nu P_{(i)}^{\mu\nu}
 \right), \label{eq:ext_susy_gen}
\end{align}
with $i=1,2$, generate transformations as $\delta^{(i)}_2\f^\mu = \{\f^\mu, \mathcal{Q}^{(i)}_2(\epsilon)\}$ and $\delta^{(i)}_2 S_\mu = \{S_\mu, \mathcal{Q}^{(i)}_2(\epsilon)\}$, where the bracket is given by \eqref{eq:twisted_Poisson_bracket}. For the Hamiltonian to have extended supersymmetry, i.e.\ that the transformations generated by $\mathcal{Q}^{(1)}_2$ or $\mathcal{Q}^{(2)}_2$ are supersymmetry transformations, the target space must be a generalized complex manifold \cite{Zabzine:2005qf} and the tensors $J_{(i)}$, $P_{(i)}$ and $L_{(i)}$ must define $H$-twisted generalized complex structures\footnote{For the definition of a $H$-twisted generalized complex structure see section \ref{sec:integrability} or \cite{Zabzine:2005qf,Bredthauer:2006hf}.} $\genJ_1$ and $\genJ_2$ as
\begin{align}
\genJ_1 = \gcgMatrix{-J_{(1)}&P_{(1)}\\L_{(1)}&J_{(1)}^t}, \;\;\;\;\;\;\;\;
\genJ_2 = \gcgMatrix{-J_{(2)}&P_{(2)}\\L_{(2)}&J_{(2)}^t}.
\end{align}
The condition that the bracket between $\mathcal{Q}^{(1)}_2$ and $\mathcal{Q}^{(2)}_2$ should give rise to the Hamiltonian \eqref{eq:Hamiltonian}, i.e.\ $\{\mathcal{Q}^{(1)}_2(\epsilon_1),\mathcal{Q}^{(2)}_2(\epsilon_2)\} = 2i\epsilon_1\epsilon_2 \mathcal{H}$,  requires that 
\begin{align}
-\genJ_1 \genJ_2 = \gcgMatrix{0&g^{-1}\\g&0}= \mathcal{G} \;\;\;\;\mbox{and}\;\;\;\;
[\genJ_1, \genJ_2]=0, \label{eq:genKahlercond}
\end{align} 
where $\mathcal{G}$ is a positive definite generalized metric on $TM\oplus TM^*$. Further, the Jacobi identity for the bracket \eqref{eq:twisted_Poisson_bracket} imply that the Hamiltonian is invariant under the transformations generated by $\mathcal{Q}^{(1)}_2$ and $\mathcal{Q}^{(2)}_2$. The conditions \eqref{eq:genKahlercond} imply that the target space is a generalized K\"{a}hler manifold, which is the main result of \cite{Bredthauer:2006hf}. In \cite{Gualtieri:2003dx} it is shown that the bi-Hermitean geometry \cite{Gates:1984nk} is equivalent to generalized K\"{a}hler geometry \cite{Gualtieri:2003dx}.

\section{T-duality of the $N=(1,1)$ supersymmetric sigma model}
Next, we review the method of performing the T-duality transformation as a gauging, and subsequent gaugefixing, of an isometry in the manifest $N=(1,1)$ sigma model, following the lines of \cite{Albertsson:2004gr}.

Consider the theory on a manifold with one compact direction, being a circle $S^1$, such that there is
an isometry described by the Killing vector $k$ along the compact direction. We thus have $\mathcal{L}_k g=0$. Further we assume that $\mathcal{L}_k b=0$. This implies that the action \eqref{eq:N11action} is invariant under the isometry transformation
\begin{align}
\delta_i(a) \Phi^\mu = a k^\mu(\Phi), \label{eq:N11_isometry}
\end{align}
which is the transformation we now want to gauge.

If $M=B\times S^1$ we can go to adapted coordinates, such that $k^\mu=(1,0,...,0)$. We then expand the action \eqref{eq:N11action} by letting $\mu=(0,m)$ and $\nu=(0,n)$, where $m,n = 1,...,D-1$. Next we gauge the isometry by introducing the gauge fields $A_\pm$ and add a Lagrange multiplier $\tilde{\Phi}^0$ to ensure that $A_\pm$ is pure gauge. The action becomes the so called parent action and is given by
\begin{align}
S= \frac{1}{2} \int\d^2\s\d\theta^+\d\theta^- \Big\{&
    \left(D_+\Phi^0 + A_+\right)\left(D_-\Phi^0 + A_-\right)g_{00}\cr
 &+ \left(D_+\Phi^0 + A_+\right) D_-\Phi^m E_{0m}\cr
 & + D_+\Phi^m \left(D_-\Phi^0 + A_-\right) E_{m0}\cr
 &+ D_+\Phi^m D_-\Phi^n E_{mn}
  - \tilde{\Phi}^0\left(D_+ A_- + D_- A_+\right)
\Big\}.\label{eq:parent_action}
\end{align}
This action is, as a consequence of \eqref{eq:spinor_relations}, invariant under the local gauge transformation $\delta \Phi^0 = \epsilon$, $\delta A_\pm =-D_\pm\epsilon$.

By variation of $\tilde{\Phi}^0$ and $A_\pm$ we find the equations of motion
\begin{align}
0 =& D_+ A_- + D_- A_+, \label{eq:eomA}\\
0 =& D_+ \tilde{\Phi}^0 + \left(D_+ \Phi^0 + A_+ \right)g_{00} + D_+\Phi^m E_{m0}, \label{eq:eom+}\\
0 =& - D_- \tilde{\Phi}^0 +\left(D_- \Phi^0 + A_- \right)g_{00} + D_-\Phi^m E_{0m}.\label{eq:eom-}
\end{align}

Equation \eqref{eq:eomA} imply that $A_\pm = D_\pm \Lambda$, for some scalar superfield $\Lambda$. Hence we find, as claimed, that $A_\pm$ is pure gauge. 

Integrating out $\tilde{\Phi}^0$, using its equations of motion, give us back the original action if we note that $\Lambda$ only gives a shift of the origin of the periodic coordinate $\Phi^0$, or equivalently if we fix the gauge $\Lambda=0$. 

If we integrate out $A_\pm$ instead, fix the gauge $\Lambda=-\Phi^0$ and identify the Lagrange multiplier $\tilde{\Phi}^0$ as the new coordinate we obtain the T-dual action 
\begin{align}
\tilde{S}= \frac{1}{2} \int\d^2\s\d\theta^+\d\theta^- \; D_+ \tilde{\Phi}^\mu D_-\tilde{\Phi}^\nu \tilde{E}_{\mu\nu}.\label{eq:Tdual-action}
\end{align}
where $\tilde{\Phi}^\mu = (\tilde{\Phi}^0, \Phi^m)$ and $\tilde{E}_{\mu\nu}=\tilde{g}_{\mu\nu}+\tilde{b}_{\mu\nu}$. The T-dual metric and $b$-field are given by the usual Buscher rules \cite{Buscher:1987qj, Alvarez:1994wj}
\begin{align}
\tilde{g}_{00} &= \frac{1}{g_{00}}, \;\;\;\;\;\;\;\;
\tilde{g}_{0m}  = -\frac{b_{0m}}{g_{00}}, \;\;\;\;\;\;\;\; 
\tilde{g}_{mn}  = g_{mn} - \frac{g_{0m}g_{0n}-b_{0m}b_{0n}}{g_{00}}\cr
\tilde{b}_{0m} &= -\frac{g_{0m}}{g_{00}}, \;\;\;\;\;\;\;\;
\tilde{b}_{mn}  = b_{mn} - \frac{g_{0m}b_{0n}-b_{0m}g_{0n}}{g_{00}}. \label{eq:Buscher_rules}
\end{align}

\section{T-duality transformation in phase space}\label{sec:T_dual_phasespace}

The easiest way to find the T-duality transformation in the phase space formulation is to perform the reduction to $N=1$ superfields of the equations of motion, \eqref{eq:eom+} and \eqref{eq:eom-}.

The original coordinate in the isometry direction, $\Phi^0$, was obtained by choosing the gauge $\Lambda=0$. This implies that we must use this gauge to find a relation between $\Phi^0$ and the T-dual coordinate $\tilde{\Phi}^0$. Using this gauge and performing the transition \eqref{eq:newcoords} to the new odd coordinates and spinor derivatives turn the equations of motion \eqref{eq:eom+} and \eqref{eq:eom-} into
\begin{align}
D_0 \tilde{\Phi}^0 = b_{0m}D_0\Phi^m - g_{0\mu} D_1\Phi^\mu, \label{eq:eomT1}\\
D_1 \tilde{\Phi}^0 = b_{0m}D_1\Phi^m - g_{0\mu} D_0\Phi^\mu. \label{eq:eomT2}
\end{align}
We now reduce these equations to $N=1$ fields. First, observe that the T-dual $N=1$ field $\tilde{S}_\mu$ must be defined with the T-dual metric, i.e.\ $\tilde{S}_\mu = \tilde{g}_{\mu\nu} D_0\tilde{\Phi}^\nu|_{\theta_0=0}$.  Secondly, note that $\Phi^m=\tilde{\Phi}^m$, i.e\ the components of $\Phi$ along the space $B$ does not change under T-duality.

Forming $\tilde{S}_0 = \tilde{g}_{0\nu}D_0\tilde{\Phi}^\nu|_{\theta^0=0}$ and $\tilde{S}_m = \tilde{g}_{m\nu}D_0\tilde{\Phi}^\nu|_{\theta^0=0}$, and using \eqref{eq:eomT1} and \eqref{eq:N=1superfields} give
\begin{align}
\tilde{S}_0 &= -D\f^0 - \frac{g_{0m}}{g_{00}}D\f^m\\
\tilde{S}_m &=  S_n - \frac{g_{0n}}{g_{00}} S_0 
                 + b_{0n} D\f^0 + b_{0n}\frac{g_{0m}}{g_{00}} D\f^m.
\end{align}
Reducing \eqref{eq:eomT2} to $N=1$ fields immediately produces the relation
\begin{align}
D\tilde{\f}^0 = - S_0 + b_{0m}D\f^m.
\end{align}
These relations define a bundle morphism between $X^*\Pi(TM\oplus T^*M)$ and $\tilde{X}^*\Pi(T\tilde{M}\oplus T^*\tilde{M})$, where $\Pi$ denotes the bundle with reversed parity on the fibers and $X^*$ and $\tilde{X}^*$ are pull backs by the bosonic component of $\f$ and $\tilde{\f}$ respectively. Thus we have found the T-duality transformation as
\begin{align}
\left\{
\begin{array}{rl}
D\tilde{\f}^0 =& - S_0 + b_{0m} D\f^m \\
D\tilde{\f}^n =&   D\f^n\\
\tilde{S}_0   =& - D\f^0 - \frac{g_{0m}}{g_{00}}D\f^m \\
\tilde{S}_n   =&   S_n - \frac{g_{0n}}{g_{00}} S_0 
                 + b_{0n} D\f^0 + b_{0n}\frac{g_{0m}}{g_{00}} D\f^m ,
\end{array}
\right.\label{eq:Tduality-transform}
\end{align}
where we have introduced spinor derivatives on the fields $\f^n$. Inverting the transformation, we find the inverse T-duality transformation as
\begin{align}
\left\{
\begin{array}{rl}
D\f^0 =& - \tilde{S}_0 - \frac{g_{0m}}{g_{00}}D\tilde{\f}^m \\
D\f^n =&   D\tilde{\f}^n\\
S_0   =& - D\tilde{\f}^0 + b_{0m}D\tilde{\f}^m \\
S_n   =&   \tilde{S}_n + b_{0n} \tilde{S}_0 
         - \frac{g_{0n}}{g_{00}} D\tilde{\f}^0 
         + \frac{g_{0n}}{g_{00}} b_{0m}D\tilde{\f}^m .
\end{array}
\right.\label{eq:Tinverse-transform}
\end{align}

Next, we transform the Hamiltonian \eqref{eq:Hamiltonian} into the T-dual Hamiltonian.
Using the transformation \eqref{eq:Tinverse-transform} in \eqref{eq:Hamiltonian} a lengthy but straightforward calculation gives
\begin{align}
\tilde{\mathcal{H}}=
\frac{1}{2}\int\d\s\d\theta \;\Big(
&   D^2\tilde{\f}^\mu D\tilde{\f}^\nu \tilde{g}_{\mu\nu}
  + \tilde{S}_\mu D\tilde{S}_\nu \tilde{g}^{\mu\nu}
  + \tilde{S}_\mu D\tilde{\f}^\rho \tilde{S}_\nu \tilde{g}^{\nu\sigma}\tilde{\Gamma}^{(0)\mu}_{\sigma\rho}\cr
& + D\tilde{\f}^\mu D\tilde{\f}^\nu \tilde{S}_\rho \tilde{H}_{\mu\nu}{}^\rho
  -\frac{1}{3} \tilde{S}_\mu \tilde{S}_\nu \tilde{S}_\rho \tilde{H}^{\mu\nu\rho} \Big),
\end{align}
where the components of the T-dual metric and $b$-field are given by \eqref{eq:Buscher_rules} and the components of the T-dual inverse metric are given by
\begin{align}
\tilde{g}^{00} = g_{00} + g^{mn}b_{0m}b_{0n}, \;\;\;\;\;\;\;\;
\tilde{g}^{0m} = g^{mn}b_{0n}, \;\;\;\;\;\;\;\;
\tilde{g}^{mn} = g^{mn}.
\end{align}
This is the correct T-dual Hamiltonian which also can be obtained from the T-dual manifest $N=(1,1)$ action \eqref{eq:Tdual-action} by going to the T-dual phase space formulation via a similar calculation to the one presented in section \ref{sec:to_phase_space}. Thus the diagram 
\begin{align}
\begin{array}{ccc}
S(g,H) & \longrightarrow & \tilde{S}(\tilde{g},\tilde{H})\\
\downarrow& & \downarrow\\
\mathcal{H}(g,H) & \longrightarrow& \tilde{\mathcal{H}}(\tilde{g},\tilde{H})
\end{array}
\end{align}
commutes.

To show that the transformation \eqref{eq:Tduality-transform} is a symplectomorphism, note that it 
transforms the integrand of \eqref{eq:N1reducedaction} as 
\begin{align}
\int \d\s\d\theta\;&
\Big\{i\left(S_\mu-b_{\mu\nu}D\f^\nu\right)\partial_0 \f^\mu - \mathcal{H} \Big\} \longrightarrow\cr
& \int \d\s\d\theta\;
   \left\{ i\left(\tilde{S}_\mu-\tilde{b}_{\mu\nu}D\tilde{\f}^\nu\right)\partial_0 \tilde{\f}^\mu - \tilde{\mathcal{H}} \right\}
 +\partial_0 \left(i \int \d\s\d\theta\;  \tilde{\f^0} D\f^0\right)
\end{align}
meaning that the T-duality transformation preserves the form of the Hamiltonian equations of motion but changes the symplectic structure and hence the Poisson bracket to their T-duals. Thus we find that the transformation behaves as expected and is a generalization of a canonical transformation, a symplectomorphism.

When $H$ is exact we can formulate the transformation \eqref{eq:Tduality-transform} as a canonical transformation by putting the $H$-field into the momenta instead of in the Poisson bracket. The generating function for the transformation can be found by noting that the phase space action \eqref{eq:N1reducedaction} defines the canonically conjugated momenta to $\f^\mu$ as
\begin{align}
P_\mu = i\left(S_\mu-b_{\mu\nu}D\f^\nu\right).
\end{align} 
In this picture the independent phase space coordinates are $\left(\f^\mu,P_\mu\right)$, which obey the standard Poisson bracket relations, but now with the untwisted Poisson bracket, obtained form \eqref{eq:twisted_Poisson_bracket} by setting $H=0$. If we next define the T-dual canonical momenta as
\begin{align}
\tilde{P}_\mu = i\left(\tilde{S}_\mu-\tilde{b}_{\mu\nu}D\tilde{\f}^\nu\right)
\end{align}
the T-duality transformation \eqref{eq:Tduality-transform} may be written as
\begin{align}
P_0   = -i D\tilde{\f}^0,\;\;\;\;\;
D\f^0 = i \tilde{P}_0,\;\;\;\;\;
\f^n  = \tilde{\f}^n, \;\;\;\;\;
P_n   = \tilde{P}_n.
\label{eq:Ttransf_canonical}
\end{align}
The generating function for this transformation is of the form $F=F(\f^0,\tilde{\f}^0,\f^n, \tilde{P}^n)$ and found to be
\begin{align}
F = \int\d\s\d\theta \left( i\tilde{\f}^0 D\f^0 + \f^n\tilde{P}_n\right).
\end{align}
The equations that generate the transformation are given by
\begin{align}
P_0         =  \frac{\overrightarrow{\delta} F}{\delta \f^0}, \;\;\;\;\;
\tilde{P}_0 = -\frac{\overrightarrow{\delta} F}{\delta \tilde{\f}^0}, \;\;\;\;\;
P_n = \frac{\overrightarrow{\delta} F}{\delta \f^n}, \;\;\;\;\;
\tilde{\f}^n = \frac{\overrightarrow{\delta} F}{\delta \tilde{P}_n}
\end{align}
with the functional derivatives being defined analogously as the ones defined in \eqref{eq:def_func_deriv}.

Now we turn to the question of the isometry. Reducing the isometry transformation \eqref{eq:N11_isometry} to $N=1$ fields and using $\mathcal{L}_k g = 0$ gives the transformation
\begin{align}
\delta_i (a) \f^\mu = a k^\mu, \;\;\;\;\;\delta_i (a) S_\mu = - a k^\nu{}_{,\mu}S_\nu \label{eq:hamiltonian_isometry}
\end{align}
where $k^\mu=k^\mu(\f)$. Since the Hamiltonian \eqref{eq:Hamiltonian} is obtained by the same reduction procedure it is invariant under this reduced isometry transformation. Requiring that a supersymmetry transformation generated by a charge like the ones in \eqref{eq:ext_susy_gen}\footnote{This supersymmetry transformation is given explicitly in \cite{Zabzine:2005qf}.} to commute with this isometry transformation we find that the $H$-twisted generalized complex structure $\genJ$, defining the transformation, must obey $\mathcal{L}_k \genJ = 0$. This is an essential property of the $H$-twisted generalized complex structure that we shall need to prove the integrability of its T-dual in section \ref{sec:integrability}.

In the above construction of the T-duality transformation we have used adapted local coordinates such that the isometry direction is the $\f^0$ direction. We now want to formulate the transformation in a covariant way. For this, note that the target space $M=B\times S^1$ is a principal bundle on which we introduce a connection $A$. Further we introduce a connection $\tilde{A}$ and a Killing vector $\tilde{k}$ on the T-dual target bundle $\tilde{M}$. We require these connections to obey $i_k A=1$ and $i_{\tilde{k}}\tilde{A}=1$. In the adapted coordinates these conditions imply $A_0=\tilde{A}_0=1$.

Using the Buscher rules \eqref{eq:Buscher_rules} we find the T-dual of the $H$-field as
\begin{align}
\tilde{H}_{0ab} &= \frac{1}{2}\left(\partial_a\left(\frac{g_{0b}}{g_{00}}\right) 
                         - \partial_b\left(\frac{g_{0a}}{g_{00}}\right)\right) \label{eq:Hdual0ab}\\
\tilde{H}_{abc} &= H_{abc} - b_{0a}\tilde{H}_{0bc} - \frac{g_{0a}}{g_{00}} H_{0bc}
                           - b_{0b}\tilde{H}_{0ca} - \frac{g_{0b}}{g_{00}} H_{0ca} 
                           - b_{0c}\tilde{H}_{0ab} - \frac{g_{0c}}{g_{00}} H_{0ab}
\end{align}
where we, in the second relation, have used $\tilde{H}_{0ab}$ as a convenient abbreviation.
We find that if the connections and 3-form field $h$ are given by
\begin{align}
&A = d\f^0 + \frac{g_{0a}}{g_{00}} d\f^a, \;\;\;\;\;\;\;\;
\tilde{A} = d\tilde{\f}^0 - b_{0a} d\tilde{\f}^a,\label{eq:connections}\\ 
&h_{0ab} = 0, \;\;\;\;\;\;\;\;
h_{abc}  = H_{abc} - \frac{2}{g_{00}} g_{0[a}H_{0|bc]} 
\end{align}
or, written in a coordinate independent way
\begin{align}
A = \frac{1}{i_k i_k g} i_k g, \;\;\;\;\;\;\;\;
\tilde{A} = \frac{1}{i_{\tilde{k}} i_{\tilde{k}} \tilde{g}} i_{\tilde{k}} \tilde{g}, \;\;\;\;\;\;\;\;
h = H - A\wedge i_k H,
\end{align}
we can write $H = d\tilde{A}\wedge A + h$ and $\tilde{H} = d A\wedge\tilde{A} + h$.
The 3-form field $h$ satisfies $i_k h=0$. Further, note that in adapted coordinates the fields $H$ and $\tilde{H}$ are independent of the isometry direction, hence they satisfy $\mathcal{L}_k H =0$ and $\mathcal{L}_{\tilde{k}} \tilde{H}=0$.
With these definitions the T-duality transformation \eqref{eq:Tduality-transform} can be written in a covariant way
\begin{align}
\left\{
\begin{array}{rl}
\tilde{A}_\mu D\tilde{\f}^\mu &= -k^\mu S_\mu \\
\tilde{k}^\mu \tilde{S}_\mu &= - A_\mu D\f^\mu\\
\big(\delta^\mu_\nu + \tilde{k}^\mu\tilde{k}^\rho \tilde{b}_{\rho\nu}\big)D\tilde{\f}^\nu 
 - \tilde{k}^\mu \tilde{k}^\rho \tilde{S}_\rho &=
\big(\delta^\mu_\nu + k^\mu k^\rho b_{\rho\nu} \big)D\f^\nu - k^\mu k^\rho S_\rho\\
\big(\delta^\nu_\mu - \tilde{k}^\nu \tilde{A}_\mu \big)\tilde{S}_\nu &= 
\big(\delta^\nu_\mu - k^\nu A_\mu \big)S_\nu .
\end{array}
\right. \label{eq:covariantTduality}
\end{align}

\section{Matrix formulation of the T-duality transformation}\label{sec:Matrix_form}
In this section we formulate the T-duality transformation in a matrix form. This enables us to find the T-dual generalized almost complex structure in this model. We give its components explicitly.

\def\tVec#1#2#3#4{\matrix{c}{#1\\#2\\#3\\#4}}
\def\tLambda#1{\tVec{D\f^0}{D\f^#1}{S_0}{S_#1}}
\def\tildetLambda#1{\tVec{D\tilde{\f}^0}{D\tilde{\f}^#1}{\tilde{S}_0}{\tilde{S}_#1}}
\def\tMatrix#1{\matrix{cccc}{#1}}
\def\Tpairing#1#2{\tMatrix{ 0&0&1&0 \\ 0&0&0&\delta^#2_#1 \\ 1&0&0&0 \\ 0&\delta^#1_#2&0&0}}
\def\Tdual#1#2{\tMatrix{0       & b_{0#2}                       & -1                      & 0           \\
                        0       & \delta^#1_#2                  & 0                       & 0           \\
                        -1      & -\frac{g_{0#2}}{g_{00}}       & 0                       & 0           \\
                        b_{0#1} & \frac{b_{0#1}g_{0#2}}{g_{00}} & -\frac{g_{0#1}}{g_{00}} & \delta^#2_#1} }
\def\invTdual#1#2{\tMatrix{0                       & -\frac{g_{0#2}}{g_{00}}       & -1      & 0            \\
                           0                       & \delta^#1_#2                  & 0       & 0            \\
                           -1                      & b_{0#2}                       & 0       & 0            \\
                           -\frac{g_{0#1}}{g_{00}} & \frac{g_{0#1}b_{0#2}}{g_{00}} & b_{0#1} & \delta^#2_#1 } }
We begin by defining a section $\Lambda$ of the pullback $X^*\Pi( TM \oplus T^*M)$ and $\tilde{\Lambda}$ a section of the T-dual space $\tilde{X}^*\Pi(T\tilde{M}\oplus T^*\tilde{M})$ as
\begin{align}
\Lambda = \tLambda{m}, \;\;\;\;\;\;\;\;
\tilde{\Lambda} = \tildetLambda{m}.
\end{align} 
Since the T-duality transformation \eqref{eq:Tduality-transform} and its inverse  \eqref{eq:Tinverse-transform} naturally act on $\Lambda$ and $\tilde{\Lambda}$ we define the bundle morphism $T: TM \oplus T^*M \longrightarrow T\tilde{M}\oplus T^*\tilde{M}$ and its inverse
${T^{-1}:T\tilde{M}\oplus T^*\tilde{M} \longrightarrow TM \oplus T^*M}$ as
\begin{align}
T = \Tdual{n}{m},
\;\;\;\;\;\;\;\;
T^{-1} = \invTdual{n}{m}. \label{eq:T-duality_matrixform}
\end{align}
With these definitions the transformations \eqref{eq:Tduality-transform} and \eqref{eq:Tinverse-transform} are written as $\tilde{\Lambda}= T\Lambda$ and $\Lambda = T^{-1}\tilde{\Lambda}$. Note that performing another T-duality transformation of the T-dual space is the same as taking the inverse of the original T-duality transformation, i.e\ $\tilde{T}=T^{-1}$, where we use the Buscher rules \eqref{eq:Buscher_rules} to find $\tilde{T}$. Hence $\tilde{T}T = \mathbbm{1}_{2d\times2d}$, as expected by T-duality.

The natural pairings $\mathcal{I}$ on $TM\oplus T^*M$ and on $T\tilde{M}\oplus T^*\tilde{M}$ are both given by
\begin{align}
\mathcal{I} = \Tpairing{m}{n}.
\end{align}
and we find that $\mathcal{I} = T^t \mathcal{I} T$. This means that $T\in O(d,d)$ and hence
leaves the inner product $\langle A, B\rangle = A^t \mathcal{I} B$ invariant.

Next we write the $H$-twisted generalized complex structure $\genJ: TM\oplus T^*M\longrightarrow TM\oplus T^*M$ and the T-dual map ${\tilde{\genJ}:T\tilde{M}\oplus T^*\tilde{M}\longrightarrow T\tilde{M}\oplus T^*\tilde{M}}$ as
\def\TgenJ#1#2{\tMatrix{-J^0_0 & -J^0_#2 & 0 & P^{0#2} \\ -J^#1_0 & -J^#1_#2 & P^{#1 0} & P^{#1#2} \\ 
                        0 & L_{0#2} & J^0_0 & J^#2_0 \\ L_{#1 0} & L_{#1#2} & J^0_#1 & J^#2_#1}}
\def\tildeTgenJ#1#2{\tMatrix{-\tilde{J}^0_0 & -\tilde{J}^0_#2 & 0 & \tilde{P}^{0#2} \\ 
                             -\tilde{J}^#1_0 & -\tilde{J}^#1_#2 & \tilde{P}^{#1 0} & \tilde{P}^{#1#2} \\ 
     0 & \tilde{L}_{0#2} & \tilde{J}^0_0 & \tilde{J}^#2_0 \\ \tilde{L}_{#1 0} & \tilde{L}_{#1#2} & \tilde{J}^0_#1 & \tilde{J}^#2_#1}}
\begin{align}
\genJ = \TgenJ{n}{m},
\;\;\;\;\;\;\;\;
\tilde{\genJ} = \tildeTgenJ{n}{m}.
\end{align}

A generator of the extended supersymmetry \eqref{eq:ext_susy_gen} defined by the $H$-twisted generalized complex structure $\genJ$ can be written in this language as \cite{Zabzine:2005qf} 
\begin{align}
\mathcal{Q}_2(\epsilon) = -\frac{1}{2}\int \d\sigma\d\theta\; \epsilon\langle\Lambda,  \genJ\Lambda\rangle. \label{eq:ext_susy_gen_wJ}
\end{align}
For the T-dual space to have extended supersymmetry the map $\tilde{\genJ}$ in the generator
\begin{align}
\tilde{\mathcal{Q}}_2(\epsilon) =& -\frac{1}{2}\int \d\sigma\d\theta\; \epsilon\langle\tilde{\Lambda}, \tilde{\genJ}\tilde{\Lambda}\rangle. \label{eq:T-dual_ext_susy_gen}
\end{align}
must be an $\tilde{H}$-twisted generalized complex structure. We now relate the generators $\mathcal{Q}_2$ and $\tilde{\mathcal{Q}}_2$ by noting that
\begin{align}
\langle \tilde{\Lambda}, \tilde{\genJ} \tilde{\Lambda}\rangle 
= \langle T \Lambda, \tilde{\genJ} T\Lambda\rangle
= \Lambda^t T^t \mathcal{I} \tilde{\genJ} T \Lambda
= \Lambda^t \mathcal{I} T^{-1} \tilde{\genJ} T \Lambda 
= \langle \Lambda, T^{-1}\tilde{\genJ} T\Lambda\rangle .
\end{align}
Hence, for $\tilde{\mathcal{Q}}_2$ to be the T-dual of the generator $\mathcal{Q}_2$ we need that ${\genJ = T^{-1}\tilde{\genJ} T}$ or equivalently ${\tilde{\genJ} = T \genJ T^{-1}}$. This implies that $\tilde{\genJ}^2=-1$ and $\tilde{\genJ}^t \mathcal{I}=-\mathcal{I}\tilde{\genJ}$, meaning that $\tilde{\genJ}$ is a generalized almost complex structure. To show that the T-dual space has the same amount of extended supersymmetry as the original space we need to show that $\tilde{\genJ}$ is integrable with respect to the $\tilde{H}$-twisted Courant bracket, defined by \eqref{eq:HCourant}, which is the subject of the next section.

The components of the T-dual generalized almost complex structure $\tilde{\genJ}=T\genJ T^{-1}$ are identified as
\begin{align}
\tilde{J}^0_0 =& - J^0_0 
                 - \frac{g_{0m}}{g_{00}} J^m_0
                 - b_{0m} P^{0m}
                 - \frac{g_{0m}}{g_{00}}b_{0n}  P^{mn} \label{eq:tildeJcomponents_1}\\
\tilde{J}^a_0 =&   P^{a0} 
                 + P^{am} \frac{g_{0m}}{g_{00}} \\
\tilde{J}^0_a =&   L_{0a} 
                 + b_{0a} J^0_0 
                 + b_{0a}\frac{g_{0m}}{g_{00}} J^m_0 
                 - \frac{g_{0a}}{g_{00}}b_{0m} J^m_0
                 + b_{0n} J^n_a
                 + b_{0a} b_{0m} P^{0m}\cr
               & + b_{0a} \frac{g_{0m}}{g_{00}}b_{0n} P^{mn}\\
\tilde{J}^b_a =&   J^b_a
                 - \frac{g_{0a}}{g_{00}} J^b_0 
                 + b_{0a} P^{0b}
                 + b_{0a} \frac{g_{0m}}{g_{00}} P^{mb}\\
\tilde{L}_{0a} =&   J^0_a
                  - \frac{g_{0a}}{g_{00}} J^0_0 
                  + \frac{g_{0n}}{g_{00}} J^n_a
                  - \frac{g_{0a}}{g_{00}} \frac{g_{0m}}{g_{00}} J^m_0 \\
\tilde{L}_{ab} =&   L_{ab}
                  - 2 b_{0[a} J^0_{b]}
                  - 2 \frac{g_{0n}}{g_{00}} b_{0[a} J^n_{b]}
                  - 2 \frac{1}{g_{00}} g_{0[a} L_{0|b]}
                  - 2 \frac{g_{0[a} b_{0|b]}}{g_{00}} J^0_0 \cr
                & - 2 \frac{g_{0[a} b_{0|b]}}{g_{00}} \frac{g_{0n}}{g_{00}} J^n_0 \\
\tilde{P}^{0a} =& - J^a_0 
                  + b_{0m} P^{ma}\\
\tilde{P}^{ab} =& P^{ab}. \label{eq:tildeJcomponents_2}
\end{align}
In these relations we use the convention $A_{[ab]}\equiv\frac{1}{2}\left(A_{ab}-A_{ba}\right)$. Note
that the above relations hold also for the untwisted case, i.e.\ when $H=0$.

\section{Integrability of a T-dual generalized complex structure}\label{sec:integrability}
In the previous section we found an explicit expression for how an $H$-twisted generalized complex structure transforms under T-duality. We also found that the T-dual object, $\tilde{\genJ}$, is a generalized almost complex structure and in this section we turn to the question of integrability. 

We begin by defining what we mean by integrability. Let $X+\xi$ and $Y+\eta$ be smooth sections of $TM\oplus T^*M$. In agreement with \cite{Zabzine:2005qf, Bredthauer:2006hf}, we define the $H$-twisted Courant bracket by
\begin{align}
\left[X+\xi, Y+\eta\right]_H &= \left[X+\xi, Y+\eta\right]_C +i_X i_Y H  \label{eq:HCourant}\\
\left[X+\xi, Y+\eta\right]_C &= [X,Y] + \mathcal{L}_X \eta - \mathcal{L}_Y \xi 
-\tsfrac{1}{2}d\left(i_X \eta -i_Y \xi\right)
\end{align}
where $[\cdot,\cdot]_C$ is the Courant bracket and $[\cdot,\cdot]$ is the standard Lie bracket on $TM$. For a generalized almost complex structure $\genJ$ to be integrable with respect to the $H$-twisted Courant bracket we require that its $+i$ eigenbundle is involutive under the $H$-twisted Courant bracket. An $H$-twisted generalized complex structure is a generalized almost complex structure that is integrable with respect to the $H$-twisted Courant bracket.

By consistency, the T-dual $\tilde{\genJ}$ has to be integrable. This follows because, since the algebra of the generators of the extended supersymmetry does not change under the symplectomorphism the T-dual generators $\tilde{\mathcal{Q}}_2^{(i)}$, of the form \eqref{eq:T-dual_ext_susy_gen}, are also supersymmetry generators. Hence, the corresponding $\tilde{\genJ}^{(i)}$'s have to be $\tilde{H}$-twisted generalized complex structures and thus, integrable with respect to the $\tilde{H}$-twisted Courant bracket. However, for clarity, we provide a proof of this statement and find that the T-dual of a $H$-twisted generalized complex structure is indeed a $\tilde{H}$-twisted generalized complex structure.

We follow a slightly modified path of that given in \cite{Cavalcanti:2005hq} to find a proof of integrability applicable to our setting. We shall call a general tensor $F$ invariant if it satisfies $\mathcal{L}_k F = 0$. The derivation is based on the formulation of T-duality in the papers \cite{Bouwknegt:2003zg,Bouwknegt:2003vb}. In this setting it is required that the fields $H$, $\tilde{H}$ and $\genJ$ are invariant, which are the properties we found in the previous sections.

The T-duality transformation of an invariant form $\rho$ on $M$ is given in terms of the connections $A$ and $\tilde{A}$ in \eqref{eq:connections}. It is given by \cite{Bouwknegt:2003zg,Bouwknegt:2003vb}
\begin{align}
\tau(\rho) = \frac{1}{2\pi}\int_{S^1} e^{A\wedge\tilde{A}}\rho, \label{eq:tau}
\end{align}
where the integration is around the T-duality circle $S^1$. Note that
$\int_{S^1} A = 2\pi$. Note also that any invariant form $\rho$ may be written as $\rho= \rho_0 +  A\rho_1$ and that the T-duality transformation of this invariant form 
is $\tau(\rho_0 + A\rho_1) = \rho_1 +\tilde{A}\rho_0$. 

Next define the differential $d_{H}=d+H\wedge$ on $M$ and $d_{\tilde{H}}=d+\tilde{H}\wedge$ on $\tilde{M}$. It then follows that
\begin{align}
\tau(d_{-H} \rho) = -d_{-\tilde{H}}\tau({\rho}) \label{eq:TdHrho}
\end{align}
for any invariant form $\rho$. In deriving this equality we use the relations $H=d\tilde{A}\wedge A +h$ and $\tilde{H}=d A\wedge\tilde{A}+h$ that we found in section \ref{sec:T_dual_phasespace}.

Next, we must define the T-duality transformation of sections of $X^*\Pi(TM\oplus T^*M)$. Let $V$ be an invariant section of $TM\oplus T^*M$, we may then write $V = Z + f\frac{\partial}{\partial A} + \xi + g A$, where $Z$ is a horizontal section with respect to the connection $A$ and $\xi$ does not have any component in the $A$-direction. The T-dual of $V$ is now given by the map
\begin{align}
\varphi\left(Z + f\frac{\partial}{\partial A} + \xi + g A\right) =
  Z - g\frac{\partial}{\partial\tilde{A}} +\xi - f\tilde{A}. \label{eq:varphi_transf}
\end{align}
Here, the transformed $Z$ is a horizontal section of $T\tilde{M}$ with respect to the connection $\tilde{A}$. We transform the section $D\f+S$ of $X^*\Pi(TM\oplus T^*M)$,
\begin{align}
\hspace*{-0.5cm}
\varphi\left(D\f+S\right) =&
 \varphi\left( \left(D\f -  A(D\f) \frac{\partial}{\partial A}\right) 
              +  A(D\f) \frac{\partial}{\partial A} 
      + \left(S_a - \frac{g_{0a}}{g_{00}}S_0\right)d\f^a + S_0 A\right)\\
=&\left(D\f - \tilde{A}(D\f) \frac{\partial}{\partial\tilde{A}}\right) 
   - S_0 \frac{\partial}{\partial\tilde{A}} 
+\left(S_a - \frac{g_{0a}}{g_{00}}S_0\right) d\tilde{\f}^a - A(D\f) \tilde{A}\\
\equiv& D\tilde{\f}^a\frac{\partial}{\partial \tilde{\f}^a} 
      + D\tilde{\f}^0 \frac{\partial}{\partial\tilde{A}}
      + \tilde{S}_a d\tilde{\f}^a + \tilde{S}_0 d\tilde{\f}^0.
\end{align}
This, together with \eqref{eq:connections}, enables the identifications of the transformed component fields. We find the T-duality transformation \eqref{eq:Tduality-transform} and hence the map \eqref{eq:varphi_transf} is indeed the correct transformation. In section \ref{sec:Matrix_form} we found that that the T-duality transformation preserves the natural pairing and thus $\varphi\in O(d,d)$.

The product of a section $V=(Y+\eta)$ of $TM\oplus T^*M$ and a form $\rho$ is defined as $V\cdot\rho = (i_Y + \eta\wedge)\rho$. 
It follows that for invariant sections and forms the following relation holds
\begin{align}
\tau(V\cdot\rho) = -\varphi(V)\cdot\tau(\rho).\label{eq:TVrho}
\end{align}

Moreover we need to know how the H-twisted Courant bracket transforms under T-duality.
It is possible to write this bracket in terms of the differential $d_H$ as \cite{Cavalcanti:2005hq}
\begin{align}
[V_1,V_2]_{H} \cdot\rho =&  \tsfrac{1}{2}V_1 \cdot V_2 \cdot d_{-H} \rho
                         + \tsfrac{1}{2}d_{-H}(V_1 \cdot V_2 \cdot \rho)\cr
                         &+ V_1 \cdot d_{-H}(V_2 \cdot \rho)
                         - V_2 \cdot d_{-H}(V_1 \cdot \rho) \label{eq:Hbracket}.
\end{align} 
Using \eqref{eq:TdHrho}, \eqref{eq:TVrho} and \eqref{eq:Hbracket} we find that
\begin{align}
\varphi([V_1,V_2]_{H}) = [\varphi(V_1),\varphi(V_2)]_{\tilde{H}}.\label{eq:Tbracket}
\end{align}
Now let $L\subset TM\oplus T^*M$ be the $+i$-eigenspace of a $H$-twisted generalized complex structure $\genJ$ then \eqref{eq:Tbracket} tells us that for any $A,B\in\varphi(L)$ the bracket $[A,B]_{\tilde{H}}\in\varphi(L)$, meaning that $\varphi(L)$ is closed under the $\tilde{H}$-twisted Courant bracket. Further, $\varphi(L)$ is maximally isotropic since $\varphi\in O(d,d)$. This implies that $\varphi(L)$ is the $+i$-eigenspace of a $\tilde{H}$-twisted generalized complex structure $\tilde{\genJ}=\varphi(\genJ)$.

Since we know from section \ref{sec:Matrix_form} how a $H$-twisted generalized complex structure transforms under T-duality we have that $\tilde{\genJ}=\varphi(\genJ)=T\genJ T^{-1}$ is integrable with respect to the $\tilde{H}$-twisted Courant bracket. We thus conclude that, given a $H$-twisted generalized complex structure $\genJ$ on $TM\oplus T^*M$ such that $\mathcal{L}_k \genJ=0$, its T-dual $\tilde{\genJ}$ is a $\tilde{H}$-twisted generalized complex structure on $T\tilde{M}\oplus T^*\tilde{M}$.

\section{T-duality and extended supersymmetry}

Above we have found that the T-dual of a $H$-twisted generalized complex structure is a $\tilde{H}$-twisted generalized complex structure. We now relate this result to the amount of extended supersymmetry. 

Since the generator $\mathcal{Q}_2$ in \eqref{eq:ext_susy_gen_wJ} of extended supersymmetry demands that the map $\genJ$ is a $H$-twisted generalized complex structure and the T-dual $\mathcal{\tilde{\genJ}}$ is integrable with respect to the $\tilde{H}$-twisted Courant bracket, if the extended supersymmetry commutes with the isometry transformation \eqref{eq:hamiltonian_isometry} we conclude that the T-dual generator $\tilde{\mathcal{Q}}_2$, defined by $\mathcal{\tilde{\genJ}}$ through \eqref{eq:T-dual_ext_susy_gen}, is a generator of extended supersymmetry in the T-dual model. Hence the amount of extended supersymmetry is preserved under T-duality if the extended supersymmetry commutes with the isometry transformation \eqref{eq:hamiltonian_isometry}.

For concreteness, we have seen that the $N=(1,1)$ sigma model \eqref{eq:N11action} requires two generators $\mathcal{Q}^{(1)}_2$ and $\mathcal{Q}^{(2)}_2$ to have extended $N=(2,2)$ supersymmetry and that this in turn requires the target space to be generalized K\"{a}hler \cite{Bredthauer:2006hf}. If one of the extended supersymmetries commutes with the isometry transformation \eqref{eq:hamiltonian_isometry}, as a consequence of \eqref{eq:genKahlercond} the other does as well. Then, the T-dual of this model has the two T-dual generators of extended supersymmetry $\tilde{\mathcal{Q}}^{(1)}_2$ and $\tilde{\mathcal{Q}}^{(2)}_2$. For these T-dual generators to obey the correct supersymmetry algebra the target space is again required to be generalized K\"{a}hler.

Further, the sigma model with $N=(4,4)$ supersymmetry is required to have a generalized Hyperk\"{a}hler manifold as target space \cite{Bredthauer:2006sz, Ezhuthachan:2006yy}. Since we have shown that when the isometry transformation \eqref{eq:hamiltonian_isometry} commutes with all the extended supersymmetries they survive dualization, the T-dual of this model also has $N=(4,4)$ supersymmetry and hence its target space also has to be generalized Hyperk\"{a}hler.

\section{Examples of T-dual spaces}

From the transformation of a $H$-twisted generalized complex structure $\genJ$, (\ref{eq:tildeJcomponents_1}--\ref{eq:tildeJcomponents_2}), it is easy to find explicit examples of T-dual manifolds. 

\subsection{The T-dual of a complex manifold}

A complex manifold is described by a generalized complex structure as \cite{Gualtieri:2003dx}
\begin{align}
\genJ = \gcgMatrix{-J&0\\0&J^t}
\end{align}
where $J$ is the complex structure. We consider this space to have a compact direction $S^1$, in which we perform the T-duality. Further we need $\mathcal{L}_k J =0$, where the $k$ specifies the direction of the $S^1$. By use of (\ref{eq:tildeJcomponents_1}--\ref{eq:tildeJcomponents_2}) the T-dual $\tilde{H}$-twisted generalized complex structure is, in adapted coordinates, found to be
\def\TComplJ#1#2{\tMatrix{-\tilde{J}^0_0 & -\tilde{J}^0_#2 & 0 & \tilde{P}^{0#2} \\ 
                             0 & -\tilde{J}^#1_#2 & \tilde{P}^{#1 0} & 0 \\ 
                             0 & \tilde{L}_{0#2} & \tilde{J}^0_0 & 0 \\ 
                           \tilde{L}_{#1 0} & \tilde{L}_{#1#2} & \tilde{J}^0_#1 & \tilde{J}^#2_#1}} 
\begin{align}
\tilde{\genJ} = \TComplJ{a}{b}
\end{align}
where the components are
\begin{align}
\tilde{J}^0_0 =& - J^0_0 
                 - \frac{g_{0m}}{g_{00}} J^m_0\\
\tilde{J}^0_a =&  b_{0a} J^0_0 
                 + b_{0a} \frac{g_{0m}}{g_{00}} J^m_0 
                 - \frac{g_{0a}}{g_{00}} b_{0m} J^m_0
                 + b_{0n} J^n_a \\
\tilde{J}^b_a =&  J^b_a
                 - \frac{g_{0a}}{g_{00}} J^b_0\\
\tilde{L}_{0a} =&   J^0_a
                  - \frac{g_{0a}}{g_{00}} J^0_0 
                  + \frac{g_{0n}}{g_{00}} J^n_a
                  - \frac{g_{0a}}{g_{00}} \frac{g_{0m}}{g_{00}} J^m_0 \\
\tilde{L}_{ab} =& - 2 b_{0[a} J^0_{b]}
                  - 2 \frac{g_{0n}}{g_{00}} b_{0[a} J^n_{b]}
                  - 2 \frac{g_{0[a} b_{0|b]}}{g_{00}} J^0_0 
                  - 2 \frac{g_{0[a} b_{0|b]}}{g_{00}} \frac{g_{0n}}{g_{00}} J^n_0 \\
\tilde{P}^{0a} =& - J^a_0 .
\end{align}
Note in particular that the T-duality transformation take the complex manifold into a manifold that is no longer complex but generalized complex. 

\subsection{The T-dual of a symplectic manifold}
The generalized complex structure of a symplectic manifold is given by \cite{Gualtieri:2003dx}
\begin{align}
\genJ = \gcgMatrix{0&-\omega^{-1}\\ \omega&0}
\end{align}
where $\omega$ is the symplectic structure. As in the previous case we consider the symplectic manifold to have a compact direction $S^1$, specified by $k$, such that $\mathcal{L}_k \omega=0$. Performing the T-duality transformation in the $S^1$ direction and using (\ref{eq:tildeJcomponents_1}--\ref{eq:tildeJcomponents_2}) the T-dual $\tilde{H}$-twisted generalized complex structure is, in adapted coordinates, given by
\def\TSymplJ#1#2{\tMatrix{-\tilde{J}^0_0 & -\tilde{J}^0_#2 & 0 & \tilde{P}^{0#2} \\ 
                          -\tilde{J}^#1_0 & -\tilde{J}^#1_#2 & \tilde{P}^{#1 0} & \tilde{P}^{#1#2} \\ 
                          0 & 0 & \tilde{J}^0_0 & \tilde{J}^#2_0 \\
                          0 & \tilde{L}_{#1#2} & \tilde{J}^0_#1 & \tilde{J}^#2_#1}} 
\begin{align}
\tilde{\genJ} = \TSymplJ{a}{b}
\end{align}
where the components are
\begin{align}
\tilde{J}^0_0 =&   b_{0m}(\omega^{-1})^{0m}
                 + \frac{g_{0m}}{g_{00}} b_{0n} (\omega^{-1})^{mn}\\
\tilde{J}^a_0 =&   (\omega^{-1})^{0a} 
                 +  \frac{g_{0m}}{g_{00}} (\omega^{-1})^{ma}\\
\tilde{J}^0_a =&   \omega_{0a} 
                 + b_{0a} b_{0m} (\omega^{-1})^{m0}
                 + b_{0a} b_{0m} \frac{g_{0n}}{g_{00}} (\omega^{-1})^{mn}\\
\tilde{J}^b_a =&   b_{0a}(\omega^{-1})^{b0}
                 + b_{0a}(\omega^{-1})^{bm}\frac{g_{0m}}{g_{00}} \\
\tilde{L}_{ab} =&  \omega_{ab}
                  + 2 \frac{1}{g_{00}} g_{0[a} \omega_{b]0} \\
\tilde{P}^{0a} =& - b_{0m}(\omega^{-1})^{ma}\\
\tilde{P}^{ab} =& - (\omega^{-1})^{ab}.
\end{align}
Similarly as for the complex case the T-dual is no longer symplectic but generalized complex.

\section{Summary and discussion}
We have presented the explicit T-duality transformation of the superfields in the phase space formulation of the manifestly $N=(1,1)$ sigma model. Further we have given the explicit T-duality transformation of a $H$-twisted generalized complex structure in this model. We have also found that when the extended supersymmetry commutes with the isometry transformation \eqref{eq:hamiltonian_isometry} the T-dual generalized almost complex structure is a $\tilde{H}$-twisted generalized complex structure, which implies that T-duality transformation preserves the amount of supersymmetry. The T-dual of the $N=(2,2)$ supersymmetric sigma model has $N=(2,2)$ supersymmetry and the target space of the T-dual theory is generalized K\"{a}hler, the T-dual of the $N=(4,4)$ model has $N=(4,4)$ supersymmetry and requires a generalized Hyperk\"{a}hler target space. We have also given two explicit examples of T-dual spaces.

To prove integrability of the T-dual generalized almost complex structure we have assumed that the extended supersymmetry commutes with the isometry transformation. This realizes the the extended supersymmetry of the T-dual sigma model in the same way as in the original model and no non-localization of the supersymmetry occurs, c.f.\ \cite{Hassan:1995je}. It would be interesting to examine in which way the extended supersymmetry in the T-dual model is realized, if realized at all, when $\mathcal{L}_k \genJ \neq 0$, i.e.\ when the extended supersymmetry does not commute with the isometry transformation \eqref{eq:hamiltonian_isometry}. 

It would further be interesting to examine if the covariant form of the T-duality transformation \eqref{eq:covariantTduality} is still a symplectomorphism transforming the Poisson bracket and the Hamiltonian to their T-duals. If this is the case it would mean that the above results are valid even for non-trivial circle bundles
\begin{align*}
\begin{array}{ccc}
M&\longleftarrow&S^1\\
\downarrow&&\\
B&&
\end{array}
\end{align*}
that is, even when an adapted coordinate system does not exist. Further, it would be interesting to generalize the results to more general torus principal bundles.

\section*{Acknowledgments}
The author would like to thank Maxim Zabzine for many enlightening discussions, for proposing the subject and for commenting on the manuscript. The author would also like to thank Ulf Lindstr\"{o}m for helpful comments, discussions and for commenting on the manuscript. Further, the author would like to thank the referee for useful comments.

\begingroup\raggedright\endgroup

\end{document}